\documentclass[11pt]{article}

\usepackage{html,amsmath,amssymb,amsfonts,epsfig,cancel,cite,longtable,caption,siunitx,array,color,booktabs,setspace,arydshln}

\usepackage[T1]{fontenc}
\usepackage{dsfont}
\usepackage{mathrsfs}
\usepackage[mathscr]{euscript}
\usepackage{calligra}

\newlength{\xtrawidth}
\newlength{\xtraheight}
\numberwithin{equation}{section}

\newcommand{\be}{\begin{equation*}}
\newcommand{\ee}{\end{equation*}}
\newcommand{\beq}{\begin{equation}}
\newcommand{\eeq}{\end{equation}}
\newcommand{\ba}{\begin{array}}
\newcommand{\ea}{\end{array}}
\newcommand{\bea}{\begin{eqnarray}}
\newcommand{\eea}{\end{eqnarray}}
\newcommand{\bean}{\begin{eqnarray*}}
\newcommand{\eean}{\end{eqnarray*}}

\newcommand{\cN}{{\cal N}}

\newcommand{\cA}{{\cal A}}
\newcommand{\cB}{{\cal B}}
\newcommand{\cC}{{\cal C}}

\newcommand{\cV}{{\cal V}}

\newcommand{\fn}{\footnotesize}

\def\cjn1{{\cA, \cC^*\otimes \wedge^j \cN^*}}
\def\bjn1{{\cA, \cB^*\otimes \wedge^j \cN^*}}
\def\vjn1{{\cA, \cV^*\otimes \wedge^j \cN^*}}
\def\cjn2{{\cA, \cC\otimes \wedge^j \cN^*}}
\def\bjn2{{\cA, \cB\otimes \wedge^j \cN^*}}
\def\vjn2{{\cA, \cV\otimes \wedge^j \cN^*}}

\begin{document}
\title{{\Large\bf Matter Field K\"ahler Metric in Heterotic String Theory\\ from Localisation}}

\vspace{1cm}

\author{
\c{S}tefan Blesneag${}^{1}$, 
Evgeny I. Buchbinder${}^{2}$, 
Andrei Constantin${}^{3}$,\\
Andre Lukas${}^{1}$,
Eran Palti${}^{4}$
}
\date{}
\maketitle
{\flushright {\vskip -6cm\noindent MPP-2018-5\\UUITP-04/18\vskip 5cm}}
\thispagestyle{empty}
\begin{center} { 
	{\it 
	${}^1$Rudolf Peierls Centre for Theoretical Physics, Oxford University,\\
       1 Keble Road, Oxford, OX1 3NP, U.K.\\[0.3cm]
	${}^2$Department of Physics M013, The University of Western Australia, \\
	35 Stirling Highway, Crawley WA 6009, Australia\\[0.3cm]
       ${}^3$Department of Physics and Astronomy, Uppsala University, \\ 
       SE-751 20, Uppsala, Sweden\\[0.3cm]
       ${}^4$Max-Planck-Institut f\"ur Physik (Werner-Heisenberg-Institut)\\ F\"ohringer Ring 6, 80805 M\"unchen, Germany}}
\end{center}

\vspace{11pt}
\begin{abstract}
\noindent
We propose an analytic method to calculate the matter field K\"ahler metric in heterotic compactifications on smooth Calabi-Yau three-folds with Abelian internal gauge fields. The matter field K\"ahler metric determines the normalisations of the $\cN=1$ chiral superfields, which enter the computation of the physical Yukawa couplings. 
We first derive the general formula for this K\"ahler metric by a dimensional reduction of the relevant supergravity theory and find that its T-moduli dependence  can be determined in general. It turns out that, due to large internal gauge flux, the remaining integrals localise around certain points on the compactification manifold and can, hence, be calculated approximately without precise knowledge of the Ricci-flat Calabi-Yau metric. In a final step, we show how this local result can be expressed in terms of the global moduli of the Calabi-Yau manifold. The method is illustrated for the family of Calabi-Yau hypersurfaces embedded in $\mathbb{P}^1\times\mathbb{P}^3$ and we obtain an explicit result for the matter field K\"ahler metric in this case.
\end{abstract}
{\hbox to 7cm{\hrulefill}}
\noindent{\fn evgeny.buchbinder@uwa.edu.au},
{\fn stefan.blesneag@wadh.ox.ac.uk},
{\fn andrei.constantin@physics.uu.se}\\
{\fn lukas@physics.ox.ac.uk},
{\fn palti@mpp.mpg.de}
\newpage
\setcounter{page}{1}
\section{Introduction}
The computation of four-dimensional Yukawa couplings from string theory is notoriously difficult, mainly because methods to compute the matter field K\"ahler metric which enters the physical Yukawa couplings are lacking. In this note we report some progress in this direction. We outline a method to calculate the matter field K\"ahler metric in the context of Calabi-Yau compactifications of the heterotic string with Abelian internal gauge fluxes.\\[3mm]
Models of particle physics derived from the $E_8\times E_8$ heterotic string can be remarkably successful in accounting for the qualitative features of the Standard Model; much progress has been made in this direction, both in the older literature~ \cite{Candelas:1985en, Greene:1986bm, Greene:1986jb, Distler:1987ee, Distler:1993mk, Kachru:1995em} and more recently~\cite{Braun:2005ux, Braun:2005bw, Braun:2005nv, Bouchard:2005ag, Blumenhagen:2006ux, Blumenhagen:2006wj, Anderson:2007nc, Anderson:2008uw, Anderson:2009mh, Braun:2009qy, Braun:2011ni, Anderson:2011ns, Anderson:2012yf, Anderson:2013xka, Buchbinder:2013dna, He:2013ofa, Buchbinder:2014qda, Buchbinder:2014sya, Buchbinder:2014qca, Constantin:2015bea, Braun:2017feb}. In fact, heterotic models with the correct spectrum of the (supersymmetric) standard model can now be obtained with relative ease and in large numbers, particularly in the context of Abelian internal gauge flux~\cite{Anderson:2011ns,Anderson:2012yf}, the case we are focusing on in this note.\\[3mm]
One of the next important steps towards realistic particle physics from string theory is to find models with the correct Yukawa couplings. The calculation of physical Yukawa coupings in string theory proceeds in three steps. First, the holomorphic Yukawa couplings, that is, the trilinear couplings in the superpotential have to be determined. As holomorphic quantities, their calculation can be accomplished either by algebraic methods~\cite{Candelas:1987se,Braun:2006me, Anderson:2009ge, Anderson:2010tc} or by methods rooted in differential geometry~\cite{Candelas:1987se,Blesneag:2015pvz, Blesneag:2016yag, Buchbinder:2016jqr}.
The second step is the calculation of the matter field K\"ahler metric which determines the field normalisation and the re-scaling required to convert the holomorphic into the physical Yukawa couplings. As a non-holomorphic quantity, the matter field K\"ahler metric is notoriously difficult to calculate since it requires knowledge of the Ricci-flat Calabi-Yau metric for which analytical expressions are not available. This technical difficulty has held up progress in calculating Yukawa couplings from string theory for a long time and it will be the focus of the present paper.\\[3mm]
The third step consists of stabilising the moduli and inserting their values into the moduli-dependent expressions for the physical Yukawa couplings to obtain actual numerical values. We will not address this step in the present paper, but rather focus on developing methods to calculate the matter field K\"ahler metric as a function of the moduli.\\[3mm]
The only class of heterotic Calabi-Yau models where an analytic expression for the matter field K\"ahler metric is known is for models with standard embedding of the spin connection into the gauge connection. In this case, the matter field K\"ahler metrics for the $(1,1)$ and $(2,1)$ matter fields are essentially given by the metrics on the corresponding moduli spaces~\cite{Candelas:1987se, Candelas:1990pi}. Recently, Candelas, de la Ossa and McOrist \cite{Candelas:2016usb} (see also Ref.~\cite{McOrist:2016cfl}) have proposed an $\alpha'$-correction of the heterotic moduli space metric, which includes bundle moduli. This information may be used to infer the K\"ahler metric of matter fields that arise from bundle moduli. However, we will not pursue this method here, since our main interest is not in bundle moduli but in the gauge matter fields which can account for the physical particles.\\[3mm]
There are two other avenues for calculating the matter field K\"ahler metric suggested by results in the literature. The first one relies on Donaldson's numerical algorithm to determine the Ricci-flat Calabi-Yau metric~\cite{Donaldson:2001, Donaldson:2005, DonaldsonNumerical} and subsequent work applying this algorithm to various explicit examples and to the numerical calculation of the Hermitian Yang-Mills connection on vector bundles~\cite{Wang:2005, Headrick:2005ch, Douglas:2006hz, Doran:2007zn, Headrick:2009jz, Douglas:2008es, Anderson:2010ke, Anderson:2011ed}. At present, this approach has not been pushed as far as numerically calculating physical Yukawa couplings. However, it appears that this is possible in principle and, while constituting a very significant computational challenge, would be very worthwhile carrying out. A disadvantage of this method is that it will only provide the Yukawa couplings at specific points in moduli space and that extracting information about their moduli dependence will be quite difficult.\\[3mm]
In this paper, we will focus on a different approach, based on localisation due to flux, which can lead to analytic results for the matter field K\"ahler metric. This method is motivated by work in F-theory~\cite{Heckman:2008qa, Font:2009gq, Conlon:2009qq, Aparicio:2011jx, Palti:2012aa} where the localisation of matter fields on the intersection curves of D7-branes and Yukawa couplings on intersections of such curves facilitates local computations of the Yukawa couplings which do not require knowledge of the Ricci-flat Calabi-Yau metric. It is not immediately obvious whether and how this approach might transfer to the heterotic case, since heterotic compactifications lack the intuitive local picture, related to intersecting D-brane models, which is available in F-theory. In this paper, we will show, using methods from differential geometry developed in Refs.~ \cite{Blesneag:2015pvz, Blesneag:2016yag, Buchbinder:2016jqr}, that localisation of wave functions can nevertheless arise in heterotic models. The underlying mechanism is, in fact, similar to the one employed in F-theory. Sufficiently large flux - in the heterotic case $E_8\times E_8$ gauge flux - leads to a localisation of wave functions which allows calculating their normalisation locally, without recourse to the Ricci-flat Calabi-Yau metric.\\[3mm] 
To carry this out explicitly we will proceed in three steps. First, we derive the general formula for the matter field K\"ahler metric for heterotic Calabi-Yau compactifications by a standard reduction of the 10-dimensional supergravity. This formula, which provides the matter field K\"ahler metric in terms of an integral over harmonic bundle valued forms is not, in itself, new (see, for example, Ref.~\cite{Lukas:1999yn}). Our re-derivation serves two purposes. First, we would like to fix conventions and factors as this will be required for an accurate calculation of the physical Yukawa couplings and, secondly, we will show explicitly how this formula for the matter field K\"ahler metric is consistent with four-dimensional $N=1$ supergravity. We observe that this consistency already determines the dependence of the matter field K\"ahler metric on the T-moduli, a result which, to our knowledge, has not been pointed out in the literature so far.\\[3mm]
The second step is to show how (Abelian) $E_8\times E_8$ gauge flux can lead to a localisation of the matter field wave functions around certain points of the Calabi-Yau manifold. We will first demonstrate this for toy examples based on line bundles on $\mathbb{P}^1$ as well as on products of projective spaces and then show that the effect generalises to Calabi-Yau manifolds. As a result, we obtain local matter field wave functions on Calabi-Yau manifolds and explicit results for their normalisation integrals.\\[3mm]
The final step is to express these results in terms of the global moduli of the Calabi-Yau manifold. We show that this can indeed be accomplished by relating global to local quantities on the Calabi-Yau manifold and by using information from four-dimensional $N=1$ supersymmetry. In this way, we can obtain explicit results for the matter field K\"ahler metric as a function of the Calabi-Yau moduli and this is carried out for the Calabi-Yau hyper-surface in $\mathbb{P}^1\times\mathbb{P}^3$. We believe this is the first time such a result for the matter field K\"ahler metric as a function of the properly defined moduli has been obtained in any geometrical string compactification, including F-theory.\\[3mm]
The plan of the paper is as follows. In the next section, we sketch the supergravity calculation which leads to the general formula for the matter field K\"ahler metric and we discuss the implications from four-dimensional $N=1$ supersymmetry. In Section~\ref{sec:proj}, we show how gauge flux leads to the localisation of matter field wave functions, starting with  toy examples on $\mathbb{P}^1$ and then generalising to products of projective spaces. Section~\ref{sec:loc} contains the local calculation of the wave function normalisation on a patch of the Calabi-Yau manifold. In Section~\ref{sec:global}, we express this result in terms of the properly defined moduli by relating global and local quantities and we obtain an explicit result for the matter field K\"ahler metric on Calabi-Yau hyper-surfaces in $\mathbb{P}^1\times\mathbb{P}^3$. We conclude in Section~\ref{sec:con}.

\section{The matter field K\"ahler metric in heterotic compactifications}\label{sec:reduction}
Our first step is to derive a general formula for the matter field K\"ahler metric, in terms of the underlying geometrical data of the Calabi-Yau manifold and the gauge bundle. The basic structure of this formula is well-known for some time, see, for example Ref.~\cite{Lukas:1999yn}, and our re-derivation here serves two purposes. Firstly, we would like to fix notations and conventions so that our result is accurate, as is required for a detailed calculation of Yukawa couplings. Secondly, we would like to explore the constraints on the matter field K\"ahler metric which arise from four-dimensional $N=1$ supergravity.\\[3mm]
Starting point is the 10-dimensional $N=1$ supergravity coupled to a 10-dimensional $E_8\times E_8$ super Yang-Mills theory. This theory contains two multiplets, namely the gravity multiplets which consists of the metric $g$, the NS two-form $B$, the dilaton $\phi$ as well as their fermionic partners, the gravitino and the dilatino, and an $E_8\times E_8$ Yang-Mills multiplet with gauge field $A$ and associated field strength $F=dA+A\wedge A$ as well as its superpartners, the gauginos. To first order in $\alpha'$ and at the two-derivative level, the bosonic part of the associated 10-dimensional action is given by
\begin{equation}
S = \frac{1}{2\kappa}\int d^{10}x\, \sqrt{-g} e^{-2\phi} \left( R + 4 \left(\partial \phi\right)^2 -\frac{1}{2} H^2 - \frac{\alpha'}{4} {\rm Tr} F^2 \right)\;,\qquad
H = dB - \frac{\alpha'}{4}\left(\omega_{\rm YM} -\omega_{\rm L}\right)\; , \label{S10}
\end{equation}
where $\kappa$ is the ten-dimensional gravitational coupling constant and $\omega_{\rm YM}$ and $\omega_{\rm L}$ are the gauge and gravitational Chern-Simons forms, respectively.\\[3mm] 
We consider the reduction of this action on a Calabi-Yau three-folds $X$, with Ricci-flat metric $g^{(6)}$ and a holomorphic bundle $V\rightarrow X$ with a connection $A^{(6)}$ that satisfies the Hermitian Yang-Mills equations, as usual. Let us introduce the K\"ahler form $J$ on $X$, related to the Ricci-flat metric $g^{(6)}$ on $X$ by $g^{(6)}_{m\bar{n}}=-iJ_{m\bar{n}}$ and a basis $J_i$, where $i=1,\ldots ,h^{1,1}(X)$, of harmonic (1,1)-forms. Then we can expand
\begin{equation}
 J=t^iJ_i\;,\qquad B=B^{(4)}+\tau^iJ_i\; ,
\end{equation}
with the K\"ahler moduli $t^i$, their axionic partners $\tau_i$ and the four-dimensional two-form $B^{(4)}$. In addition, we have the zero mode $\phi^{(4)}$ of the 10-dimensional dilaton $\phi$ as well as complex structure moduli $Z^a$, where $a=1,\ldots ,h^{2,1}(X)$. It is well-known that, in the absence of matter fields, these bosonic fields fit into four-dimensional $N=1$ chiral multiplets as
\begin{equation}
 S={\cal V}e^{-2\phi^{(4)}}+i\sigma\;,\qquad T^i=t^i+i\tau^i\;, \label{sfdef0}
\end{equation}
with the volume ${\cal V}$ of $X$ and the dual $\sigma$ of the four-dimensional two-form $B^{(4)}$. We note that the Calabi-Yau volume can be written as
\begin{equation}
 {\cal V}=\int_Xd^6x\,\sqrt{g^{(6)}}=\frac{1}{6}{\cal K}\;,\qquad {\cal K}=d_{ijk}t^it^jt^k\;,\qquad d_{ijk}=\int_X J_i\wedge J_j\wedge J_k\; , \label{Vdef}
\end{equation}
where $d_{ijk}$ are the triple intersection numbers of $X$. Further, the K\"ahler moduli space metric takes the form
\begin{equation}
{\cal G}_{ij}=-\frac{1}{4}\frac{\partial^2}{\partial t^i\partial t^j}\ln\kappa=-\frac{3}{2}\left(\frac{{\cal K}_{ij}}{{\cal K}}-\frac{3}{2}\frac{{\cal K}_i{\cal K}_j}{{\cal K}^2}\right)\;,
\end{equation}
where ${\cal K}_i=d_{ijk}t^jt^k$ and ${\cal K}_{ij}=d_{ijk}t^k$.  The complex structure moduli $Z^a$ each form the bosonic part of an $N=1$ chiral multiplet which we denote by the same name. \\[3mm]
In addition, there are matter fields $C^I$ which arise from expanding the gauge field as
\begin{equation}
 A=A^{(6)}+\nu_I C^I\; , \label{Aexp}
\end{equation}
where $\nu_I$ are harmonic one-forms which take values in the bundle $V$. It is important to stress that the correct matter field metric 
has to be computed relative to {\em harmonic} forms $\nu_I$ and this is, in fact, how the dependence on the Ricci-flat metric and the Hermitian Yang-Mills connection comes about.
The fields $C^I$ each form the bosonic part of an $N=1$ chiral supermultiplet. It is known that the definition of the $T^i$ superfields in Eq.~\eqref{sfdef0} has to be adjusted in the presence of these matter fields. In the universal case~with only one T-modulus and one matter field $C$, the required correction to Eq.~\eqref{sfdef0} has been found to be proportional to $|C|^2$ (see, for example, Ref.~\cite{Lukas:1997fg}). For our general case, we, therefore start by modifying the definition of the T-moduli in Eq.~\eqref{sfdef0} by writing
\begin{equation} 
 T^i=t^i+i\tau^i+\Gamma^i_{IJ}C^I\bar{C}^J\;,  \label{tdef}
\end{equation}
where $\Gamma^i_{IJ}$ is a set of (potentially moduli-dependent) coefficients to be determined~\footnote{The dilaton superfield $S$ receives a similar correction in the presence of matter fields~\cite{Lukas:1997fg}  but this arises at one-loop level and will not be of relevance here.} . To our knowledge, no general expression for $\Gamma^i_{IJ}$ has been obtained in the literature so far.\\[3mm]
The kinetic terms of the above superfields derive from a K\"ahler potential of the general form
\begin{equation}
 K= -\log(S+\bar{S})+K_{\rm cs}-\log\left(d_{ijk}(T^i+\bar{T}^i)(T^j+\bar{T}^j)(T^k+\bar{T}^k)\right)+G_{IJ}C^I\bar{C}^J\; , \label{K}
\end{equation} 
where $K_{\rm cs}$ is the K\"ahler potential for the complex structure moduli $Z^a$ whose explicit form is well-known but is not relevant to our present discussion and $G_{IJ}$ is the (moduli-dependent) matter field K\"ahler metric we would like to determine. The general task is now to compute the kinetic terms which result from this K\"ahler potential, insert the definitions of $S$ in Eq.~\eqref{sfdef0} and of $T^i$ in Eq.~\eqref{tdef} and compare the result with what has been obtained from the reduction of the 10-dimensional action~\eqref{S10}. This comparison should lead to explicit expressions for $G_{IJ}$ and $\Gamma_{IJ}^i$. \\[3mm]
A quick look at the K\"ahler potential~\eqref{K} shows that achieving this match is by no means a trivial matter. The matter field K\"ahler metric $G_{IJ}$ depends on the T-moduli and, hence, the kinetic terms from~\eqref{K} can be expected to include cross terms of the form $\partial_\mu t^i\partial^\mu C^I$. However, such cross terms can clearly not arise from the dimensional reduction of the 10-dimensional action~\eqref{S10} and, hence, there must be non-trivial cancellations which involve the derivatives of $G_{IJ}$ and $\Gamma_{IJ}^i$. We find that this issue can be resolved and indeed a complete match between the reduced 10-dimensional action~\eqref{S10} and the four-dimensional K\"ahler potential~\eqref{K} can be achieved provided the following three requirements are satisfied.
\begin{itemize}
\item The coefficients $\Gamma_{IJ}^i$ which appear in the definition~\eqref{tdef} of the $T^I$ superfields are given by
\begin{equation}
 \Gamma_{IJ}^i=-\frac{1}{2}{\cal G}^{ij}\frac{\partial G_{IJ}}{\partial \bar{T}^j}\;, \label{gammares}
\end{equation}
where ${\cal G}^{ij}$ is the inverse of the K\"ahler moduli space metric ${\cal G}_{ij}$. 
\item The matter field K\"ahler metric is given by
\begin{equation}
 G_{IJ}=\frac{1}{2{\cal V}}\int_X \nu_I\wedge \bar{\star}_V ({\nu}_J)\; , \label{G0}
\end{equation} 
where $\bar{\star}_V$ refers to a Hodge dual combined with a complex conjugation and an action of the hermitian bundle metric on $V$. 
\item Since the Hodge dual on a Calabi-Yau manifold acting on a $(1,0)$ form $\rho$ can be carried out as $\star\rho =-\frac{i}{2}J\wedge J\wedge \rho$ the result~\eqref{G0} for the matter field K\"ahler metric can be re-written as
\begin{equation}
 G_{IJ}=-\frac{3it^it^j}{2{\cal K}}\Lambda_{ijIJ}\;,\qquad \Lambda_{ijIJ}=\int_X J_i\wedge J_j\wedge \nu_I\wedge (H\bar{\nu}_J)\; ,\label{G1}
\end{equation}
where $H$ is the hermitian bundle metric on $V$. The final requirement for a match between the dimensionally reduced 10-dimensional and the four-dimensional theory~\eqref{K} can then be stated by saying that the above integrals $\Lambda_{ijIJ}$ do not explicitly depend on the K\"ahler moduli $t^i$. 
\end{itemize}
 The above result means that the K\"ahler moduli dependence of the matter field metric is completely determined as indicated in the first equation~\eqref{G1}, while the remaining integrals $\Lambda_{ijIJ}$ are $t^i$-independent but can still be functions of the complex structure moduli. To our knowledge this is a new result which is of considerable relevance for the structure of the matter field K\"ahler metric and the physical Yukawa couplings. Note that the $t^i$ dependence of $G_{IJ}$ in Eq.~\eqref{G1} is homogeneous of degree $-1$, as expected on general grounds.\\[3mm]
 It is worth noting that the K\"ahler potential~\eqref{K} with the matter field K\"ahler metric as given in Eq.~\eqref{G1} can, alternatively, also be written in the form
 \begin{equation}\begin{array}{lll}
   K&=& -\log(S+\bar{S})+K_{\rm cs}-\log\left(d_{ijk}(T^i+\bar{T}^i-\gamma^i)(T^j+\bar{T}^j-\gamma^j)(T^k+\bar{T}^k-\gamma^k)\right)\\
   \gamma^i&=&2\,\Gamma^i_{IJ}C^I\bar{C}^J\; ,
 \end{array}  \label{K1}
\end{equation}
provided that terms of higher than quadratic order in the matter fields $C^I$ are neglected. This can be seen by expanding the logarithm in Eq.~\eqref{K1} to leading order in $\gamma^i$ and by using $\frac{3{\cal K}_i}{\cal K}\Gamma^i_{IJ}=G_{IJ}$. (The latter identity follows from ${\cal G}^{ij}\frac{3{\cal K}_j}{4{\cal K}}=t^i$, the fact that $G_{IJ}$ is homogeneous of degree $-1$ in $t^i$ and the result~\eqref{gammares} for $\Gamma_{IJ}^i$). This form of the K\"ahler potential, together with the definition~\eqref{tdef} of the fields $T^i$, means that, in terms of the underlying geometrical K\"ahler moduli $t^i$, the dependence on the matter fields $C^I$ cancels.  Indeed, inserting the definition~\eqref{tdef} of the $T^i$ moduli into Eq.~\eqref{K1} turns the last logarithm into $-\ln(8{\cal K})$. That this part of the K\"ahler potential can be written as the negative logarithm of the Calabi-Yau volume is in fact expected and provides a check of our calculation. 

\section{Localisation of matter field wave functions on projective spaces} \label{sec:proj}
As a warm-up, we first discuss wave function normalisation on $\mathbb{P}^n$ and products of projective spaces, beginning with the simplest case of $\mathbb{P}^1$. (For a related discussion, in the context of F-theory, see Ref.~\cite{Palti:2012aa}.) In doing so we have two basic motivations in mind. First, considering projective space and $\mathbb{P}^1$ in particular provides us with a toy model for the actual Calabi-Yau case which we will tackle later. From this point of view, the following discussion will provide some intuition as to when wave function localisation occurs and when it leads to a good approximation for the normalisation integrals. On the other hand, projective spaces and their products provide the ambient spaces for the Calabi-Yau manifolds of interest and, hence, this chapter will be setting up some of the requisite notation and results we will be using later.

\subsection{Wave functions on $\mathbb{P}^1$}
Homogeneous coordinates on $\mathbb{P}^1$ are denoted by $x_0$, $x_1$, the affine coordinates on the patch $\{x_0\neq 0\}$ by $z=x_1/x_0$ and we also define $\kappa=1+|z|^2$. For simplicity, we will write all quantities in terms of the affine coordinate $z$ and we will ensure they are globally well-defined by demanding the correct transformation property under the transition $z\rightarrow 1/z$. In terms of $z$, the standard Fubini-Study K\"ahler potential and K\"ahler form can be written as
\begin{equation}
 K=\frac{i}{2\pi}\ln\kappa\;,\qquad J=\partial\bar{\partial}K=\frac{i}{2\pi\kappa^2}dz\wedge d\bar{z}\; . \label{JP1}
\end{equation}
Here, $J$ has the standard normalisation, that is, $\int_{\mathbb{P}^1}J=1$.  The associated Fubini-Study metric is K\"ahler-Einstein and, hence, the closest analogue of a Ricci-flat Calabi-Yau metric we can hope for on $\mathbb{P}^1$.\\[3mm]
We are interested in line bundles $L={\cal O}_{\mathbb{P}^1}(k)$ on $\mathbb{P}^1$ with first Chern class $c_1(L)=kJ$ on which we introduce a hermitian structure with the bundle metric and the associated (Chern) connection and field strength given by
\begin{equation}
 H=\kappa^{-k}\; ,\qquad A=\partial\ln\bar{H}=-\frac{k\bar{z}}{\kappa}dz\;,\qquad F=dA=\bar{\partial}\partial\ln\bar{H}=-2\pi ik J\; .
\end{equation} 
The analogue of the harmonic forms $\nu_I$ in Eq.~\eqref{Aexp} associated to matter fields are harmonic $L$-valued forms $\alpha$, that is, forms satisfying the equations
\begin{equation}
 \bar{\partial}\alpha=0\;,\qquad \partial(\bar{H}\star\alpha)=0\; , \label{harmP1}
\end{equation}
where the Hodge star is taken with respect to the Fubini-Study metric. We would like to compute their normalisation integrals
\begin{equation}
 \langle\alpha,\beta\rangle=\int_{\mathbb{P}^1}\alpha\wedge\star (H\bar{\beta})\; , \label{intP1}
\end{equation}
 the analogue of the matter field K\"ahler metric~\eqref{G0}. These harmonic forms are in one-to-one correspondence with the bundle cohomologies $H^p(\mathbb{P}^1,L)$ and, depending on the value of $k$, we should distinguish three case.
\begin{itemize}
 \item $k\geq 0$: In this case, the only non-vanishing cohomology of $L$ is $h^0(\mathbb{P}^1,L)=k+1$, so that the relevant harmonic forms $\alpha$ are $L$-valued zero forms. The relevant solutions to Eqs.~\eqref{harmP1} are explicitly given by the degree $k$ polynomials in $z$.
\item $k=-1$: In this case, all cohomologies of $L$ vanish so there are no harmonic forms.
\item $k\leq -2$: In this case, the only non-vanishing cohomology of $L$ is $h^1(\mathbb{P}^1,L)=-k-1$ and the corresponding $L$-valued $(0,1)$-forms which solve Eqs.~\eqref{harmP1} can be written as $\alpha=\kappa^k h(\bar{z})d\bar{z}$, where $h$ is a polynomial of degree $-k-2$ in $\bar{z}$. In the following, it is useful to work with the monomial basis
\begin{equation}
 \alpha_q=\kappa^k\bar{z}^qd\bar{z}\;,\qquad q=0,\ldots ,-k-2 \label{abasis}
\end{equation}
 for these forms.
\end{itemize}
Given that the forms $\nu_I$ which appear in the actual reduction~\eqref{Aexp} are $(0,1)$-forms the most relevant case is the last one for $k\leq -2$. In this case, inserting the forms~\eqref{abasis} into the the normalisation integral~\eqref{intP1} leads to
\begin{equation}
 \langle\alpha_q,\alpha_p\rangle=-i\int_{\mathbb{C}}z^q\bar{z}^p\kappa^kdz\wedge d\bar{z}=\frac{2\pi\, q!}{(-k-1)\cdots (-k-1-q)}\delta_{qp}\; . 
\end{equation} 
In physical terminology, the integer $k$ quantifies the flux and the integer $q$ labels the families of matter fields. It is clear that the above integrals receive their main contribution from a patch near the affine origin $z\simeq 0$, provided that the flux $|k|$ is sufficiently large and the family number $q$ is sufficiently small. In this case, it seems that the above integrals can be approximately evaluated locally near $z\simeq 0$, by using the flat metric instead of the Fubini-Study metric as well as the corresponding flat counterparts of the bundle metric and the harmonic forms. Formally, these flat space quantities can be obtained from the exact ones by setting $\kappa$ to one in the expression~\eqref{JP1} for the K\"ahler form and by the replacement $\kappa^k\rightarrow e^{k|z|^2}$ in the other quantities. That is, we use the replacements
\begin{equation}
 J=\frac{i}{2\pi\kappa}dz\wedge d\bar{z}\rightarrow \frac{i}{2\pi}dz\wedge d\bar{z}\;,\qquad H=\kappa^{-k}\rightarrow e^{-k|z|^2}\;,\qquad \alpha_q=\kappa^k\bar{z}^qd\bar{z}\rightarrow e^{k|z|^2}\bar{z}^qd\bar{z}\; .
\end{equation} 
to work out the local version of the normalisation integrals which leads to
\begin{equation}
  \langle\alpha_q,\alpha_p\rangle_{\rm loc}=-i\int_{\mathbb{C}}z^q\bar{z}^pe^{k|z|^2}dz\wedge d\bar{z}=\frac{2\pi\,q!}{(-k-1)^{q+1}}\delta_{qp}\; .
\end{equation}  
For the ratio of local to exact normalisation this implies
\begin{equation}
 \frac{ \langle\alpha_q,\alpha_q\rangle_{\rm loc}}{ \langle\alpha_q,\alpha_q\rangle}=\frac{(-k-2)\cdots (-k-2-q)}{(-k-1)^{q+1}}=1-{\cal O}\left(\frac{q^2}{-k-1}\right)\; .
\end{equation} 
Hence, as long as the flux $|k|$ is sufficiently large and the family number satisfies $q^2\ll |k|$ the local versions of these integrals do indeed provide a good approximation. It is worth noting that a transformation $z\rightarrow 1/z$ to the other standard coordinate patch of $\mathbb{P}^1$ transforms the monomial basis forms $\alpha_q$ into forms of the same type but with the family number changing as $q\rightarrow (-k-1)-q$. This means that families with a large family number $q$ close to $-k-1$ in the patch $\{x_0\neq 0\}$ acquire a small family number when transformed to the patch $\{x_1\neq 0\}$ and, hence, localise at the affine origin of this patch, that is near $z=\infty$. From this point of view it is not surprising that families with large $q$ in the patch $\{x_0\neq 0\}$ cannot be dealt with by a local calculation near $z\simeq 0$. Instead, for such modes, we can carry out a local calculation analogous to the above one but near the affine origin of the patch $\{x_1\neq 0\}$.\\[3mm]
In summary, the harmonic bundle valued $(0,1)$ forms for $L={\cal O}_{\mathbb{P}^1}(k)$, where $k\leq -2$, are given by $\alpha_q$ as in Eq.~\eqref{abasis}. For sufficiently large flux $|k|$ the modes with small family number $q$ localise near the affine origin of the path $\{x_0\neq 0\}$, that is at $z\simeq 0$ and their normalisation can be obtained from a local calculation near this point. The modes with large family number $q$ localise near the affine origin of the other path $\{x_1\neq 0\}$, that is, near $z=\infty$ and their normalisation can be obtained by a similar local calculation around this point. 

\subsection{Wave functions on products of projective spaces}\label{sec:P1P3}
The previous discussion for line bundles on $\mathbb{P}^1$ can be straightforwardly generalised to line bundles on arbitrary products of projective spaces. For the sake of keeping notation simple, we will now illustrate this for the case of ${\cal A}=\mathbb{P}^1\times \mathbb{P}^3$ which is, in fact, the ambient space of the Calabi-Yau manifold on which we focus later. The situation for general products of projective spaces is easily inferred from this discussion.\\[3mm]
Homogeneous coordinates on ${\cal A}=\mathbb{P}^1\times \mathbb{P}^3$ are denoted by $x_0,x_1$ for the $\mathbb{P}^1$ factor and by $y_0,y_1,y_2,y_3$ for $\mathbb{P}^2$. The associated affine coordinates on the patch $\{x_0\neq 0,\;y_0\neq 0\}$ are $z_1=x_1/x_0$ and $z_{\alpha+1}=y_\alpha/y_0$ for $\alpha=1,2,3$ and we define $\kappa_1=1+|z_1|^2$ and $\kappa_2=1+\sum_{\alpha=2}^4|z_\alpha|^2$. The Fubini-Study K\"ahler forms for the two projective factors are~\footnote{From now on we will denote quantities defined on the ``ambient space" ${\cal A}$ by a hat in order to distinguish them from their Calabi-Yau counterparts to be introduced later.}
\begin{equation}
  \hat{J}_1=\frac{i}{2\pi}\partial\bar{\partial}\log\kappa_1=\frac{i}{2\pi\kappa_1^2}dz_1\wedge d\bar{z}_1\; ,\qquad
  \hat{J}_2=\frac{i}{2\pi}\partial\bar{\partial}\log\kappa_2=\frac{i}{2\pi\kappa_2^2}\sum_{\alpha,\beta=2}^4\left(\kappa_2\delta_{\alpha\beta}-\bar{z}_\alpha z_\beta\right)dz_\alpha\wedge d\bar{z}_\beta\; ,
\end{equation} 
and, more generally, we can introduce the K\"ahler forms
\begin{equation}
 \hat{J}=t_1\hat{J}_1+t_2\hat{J}_2\; ,
\end{equation}
with K\"ahler parameters $t_1>0$, $t_2>0$ on ${\cal A}$. Line bundles $\hat{L}={\cal O}_{\cal A}(k_1,k_2)$ with first Chern class $c_1(\hat{L})=k_1\hat{J}_1+k_2\hat{J}_2$ can be equipped with the hermitian bundle metric
\begin{equation}
 \hat{H}=\kappa_1^{-k_1}\kappa_2^{-k_2}\quad\Rightarrow\quad \hat{F}=\bar{\partial}\partial\ln\bar{H}=-2\pi i(k_1\hat{J}_1+k_2\hat{J}_2)\; .\label{Hhat}
\end{equation}
Specifically, we are interested in those line bundles $\hat{L}$ with a non-vanishing first cohomology which are precisely those with $k_1\leq -2$ and $k_2\geq 0$. In these cases
\begin{equation}
 h^1({\cal A},{\cal O}_{\cal A}(k_1,k_2))=(-k_1-1)\frac{(k_2+3)(k_2+2)(k_2+1)}{6} \label{h1A}
\end{equation} 
and a basis for the associated harmonic $\hat{L}$-valued $(0,1)$ forms is provided by
\begin{equation}
 \hat{\nu}_{\bf q}=\kappa_1^{k_1}\bar{z}_1^{\hat{q}_1}z_2^{\hat{q}_2}z_3^{\hat{q}_3}z_4^{\hat{q}_4}d\bar{z}_1\; , \label{nuhat}
\end{equation} 
where $\hat{\bf q}=(\hat{q}_1,\hat{q}_2,\hat{q}_3,\hat{q}_4)$ is a positive integer vector which labels the families and whose entries are constrained by $\hat{q}_1=0,\ldots ,-k_1-2$ and $\hat{q}_2+\hat{q}_3+\hat{q}_4\leq k_2$. Given these quantities, the integrand of the normalisation integral is proportional to
\begin{equation}
 \hat{\nu}_{\hat{\bf q}}\wedge\star (\hat{H}\bar{\hat{\nu}}_{\hat{{\bf q}}})\sim \kappa_1^{k_1}\kappa_2^{-k_2}\prod_{\alpha=1}^4|z_\alpha|^{2\hat{q}_\alpha}\; .
\end{equation} 
Hence, provided the fluxes $|k_1|$ and $k_2$ are sufficiently large and the family numbers $q_\alpha$ sufficiently small, we expect localisation on a patch $\hat{U}$ around the affine origin $z_\alpha\simeq 0$. In this case, we can again work with the flat limit where the above quantities turn into
\begin{equation}\begin{array}{lllllllllll}
 \hat{J}_1&\rightarrow&\frac{i}{2\pi}dz_1\wedge d\bar{z}_1&& \hat{J}_2&\rightarrow&\frac{i}{2\pi}\sum_{\alpha=2}^4dz_\alpha\wedge d\bar{z}_\alpha&&
 \hat{J}&\rightarrow&t_1\hat{J}_1+t_2\hat{J}_2\\[2mm]
 \hat{H}&\rightarrow&e^{-k_1|z_1|^2-k_2\sum_{\alpha=2}^4|z_\alpha|^2}&&\hat{\nu}_{\bf q}&\rightarrow&e^{k_1|z_1|^2}\bar{z}_1^{\hat{q}_1}z_2^{\hat{q}_2}z_3^{\hat{q}_3}z_4^{\hat{q}_4}d\bar{z}_1\; .
\end{array} \label{P1P3loc}
\end{equation}
A few general conclusions can be drawn from this. First, localisation near a point in ${\cal A}$ does require all fluxes $|k_i|$ to be large. If one of the fluxes is not large then localisation will happen near a higher-dimensional variety in ${\cal A}$. For example, if $|k_1|$ is not large then the wave function will localise near $\mathbb{P}^1$ times a point in $\mathbb{P}^3$. We note that such a partial localisation may actually be sufficient when we come to discuss Calabi-Yau manifolds embedded in ${\cal A}$. For example, localisation near a curve in ${\cal A}$ will typically lead to localisation near a point on a Calabi-Yau hyper-surface embedded in ${\cal A}$. Secondly, provided all $|k_i|$ are indeed large, localisation on $\hat{U}$ near the affine origin $z_\alpha\simeq 0$, for $\alpha=1,2,3,4$, requires all $\hat{q}_\alpha$ to be sufficiently small. If a certain $\hat{q}_\alpha$ is large localisation may still arise near another point in ${\cal A}$. For example, if $\hat{q}_1$ is large while the other $\hat{q}_\alpha$ are small, then localisation occurs near $z_1=\infty$, $z_2=z_3=z_4=0$.

\section{A local Calabi-Yau calculation} \label{sec:loc}
So far, we have approached the problem of computing wave function normalisations on Calabi-Yau manifolds from the viewpoint of the prospective ambient embedding spaces. In this section, we will take the complementary point of view and carry out a local calculation on a Calabi-Yau manifold. In the next section, we will show how to connect this local Calabi-Yau calculation with the ambient space point of view in order to obtain results as functions of globally defined moduli.\\[3mm]
We start with a Calabi-Yau three-fold $X$ and a line bundle $L\rightarrow X$ with a non-vanishing first cohomology and associated $L$-valued harmonic $(0,1)$ forms. Our goal is to determine the normalisation of these harmonic forms by a local calculation, assuming, at this stage, that localisation indeed occurs. To do this, we focus on a patch $U\subset X$ with local complex coordinates $Z_a$, where $a=1,2,3$, chosen such that the K\"ahler form $J$, associated to the Ricci-flat Calabi-Yau metric, is locally on $U$ well approximated by~\footnote{We will denote local quantities, defined on the patch $U$, by script symbols.}
\begin{equation}
 {\cal J}=\frac{i}{2\pi}\sum_{a=1}^3\beta_adZ_a\wedge d\bar{Z}_a\; , \label{Jloc0}
\end{equation} 
where the $\beta_a$ are positive constants. (It is, of course, possible to set $\beta_a$ equal to one by further coordinate re-definitions but, for later purposes, we find it useful to keep these explicitly.)
On $U$, we can approximate the hermitian bundle metric $H$ and the associated field strength $F$ of $L$ by
\begin{equation}
  {\cal H}=e^{-\sum_{a=1}^3K_a|Z_a|^2}\quad\Rightarrow\quad {\cal F}=\bar{\partial}\partial\ln {\cal H}=\sum_{a=1}^3K_adZ_a\wedge d\bar{Z}_a\; , \label{Hloc0}
\end{equation}  
where $K_a$ are constants which will ultimately become functions of the Calabi-Yau moduli. The Hermitian Yang-Mills equation, $J\wedge J\wedge F=0$, should be satisfied locally which leads to
\begin{equation}
 {\cal J}\wedge{\cal J}\wedge{\cal F}=0\quad\Leftrightarrow\quad \beta_1\beta_2K_3+\beta_1\beta_3K_2+\beta_2\beta_3K_1=0\; .
\end{equation}
The resulting equation for the $K_a$ will translate into a constraint on the Calabi-Yau moduli in a way that will become more explicit later. For now we should note that it implies not all $K_a$ can have the same sign (given that the $\beta_a$ need to be positive). Consider harmonic $(0,1)$-forms $v\in H^1(X,L)$. On $U$ they are approximated by $(0,1)$-forms ${\mathcal \nu}$ which must satisfy the local version of the harmonic equations
\begin{equation}
 \bar{\partial}\nu=0\;,\qquad {\cal J}\wedge{\cal J}\wedge\partial({\cal H}\nu)=0\; .
\end{equation}
In analogy with the projective case, specifically Eq.~\eqref{P1P3loc}, we assume that $K_1<0$ and $K_2,K_3>0$. Wether these sign choices are actually realised cannot be checked locally but requires making contact with the global picture - we will come back to this later. If they are, potentially localising solutions to these equations are of the form $\nu=e^{K_1|Z_1|^2}P(\bar{Z}_1,Z_2,Z_3)d\bar{Z}_1$, where $P$ is an arbitrary function of the variables indicated. Localisation of these solution still depends on the precise form of the function $P$ which cannot be determined from a local calculation. We will return to this issue in the next section when we discuss the relation to the global picture. For now, we take a practical approach and work with a monomial basis of solutions given by
\begin{equation}
 \nu_{\bf q}=e^{K_1|Z_1|^2}\bar{Z}_1^{q_1}Z_2^{q_2}Z_3^{q_3}d\bar{Z}_1\; , \label{numon}
\end{equation}
where ${\bf q}=(q_1,q_2,q_3)$ is a vector with non-negative integers. The normalisation of these monomial solutions can be explicitly computed and is given by
\begin{eqnarray} 
 M_{{\bf q},{\bf p}}&:=&\langle\nu_{\bf q},\nu_{\bf p}\rangle_{\rm loc}=\int_U\nu_{\bf q}\wedge\star({\cal H}\bar{\nu}_{\bf p})=\frac{i}{2}\delta_{{\bf q},{\bf p}}\int_U{\cal J}\wedge{\cal J}\wedge\nu_{\bf q}\wedge (H\bar{\nu}_{\bf q})\nonumber\\
 &\simeq&\frac{i}{4\pi^2}\beta_2\beta_3\delta_{{\bf q},{\bf p}}\prod_{a=1}^3\int_{\mathbb{C}}dZ_a\wedge d\bar{Z}_a|Z_a|^{2q_a}e^{-|K_a||Z_a|^2} 
  \end{eqnarray}
 After performing the integration we find for the locally-computed normalisation
 \begin{equation}
 M_{{\bf q},{\bf p}}=\langle\nu_{\bf q},\nu_{\bf p}\rangle_{\rm loc}  =2\pi\beta_2\beta_3\delta_{{\bf q},{\bf p}}\prod_{a=1}^3q_a!\,|K_a|^{-q_a-1} \; . \label{locres}
\end{equation} 
 The appearance of the exponential in each of the integrals in the second line indicates that there is indeed a chance for localisation to occur. However, the validity and practical usefulness of this result depends on a number of factors which are impossible to determine in the local picture. First of all, we should indeed have $K_1<0$ and $K_2,K_3>0$ for localisation to happen, but these conditions can only be verified by relating to the global picture. Secondly, families are defined as cohomology classes in $H^1(X,L)$ and at this stage it is not clear precisely how these relate to the monomial basis forms~\eqref{numon}. The above calculation shows that the smaller the integers in ${\bf q}=(q_1,q_2,q_3)$ the better the localisation and this ties in with the result on projective spaces in the previous section. Finding the relation between the elements of $H^1(X,L)$ and the local basis forms $\nu_{\bf q}$ is, therefore, crucial in deciding the validity and accuracy of the approximation for the physical families. Finally, we would like to express the local result~\eqref{locres} in term of the properly defined global Calabi-Yau moduli. We will now address these issues by relating the above local calculation to the full Calabi-Yau manifold.

\section{Relating local and global quantities}\label{sec:global}
We will start by relating the local quantities which have entered the previous calculation to global quantities on the Calabi-Yau manifold, starting with the K\"ahler form and the connection on the bundle and then proceeding to bundle-valued forms. This will allows us to express the result~\eqref{locres} for the wave function normalisation in terms of properly defined moduli. 

\subsection{K\"ahler form and connection}
We begin, somewhat generally, with a Calabi-Yau three-fold $X$, a basis $J_i$, where $i=1,\ldots ,h^{1,1}(X)$ of its second cohomology and K\"ahler forms
\begin{equation}
 J=\sum_i t^iJ_i \label{Jexp}
\end{equation}
with the K\"ahler moduli ${\bf t}=(t^i)$ restricted to the K\"ahler cone. Further, we assume that all the forms $J_i$, and, hence, $J$ are chosen to be harmonic relative to the Ricci-flat metric on $X$ specified by the K\"ahler class $[J]$. Note that, despite what Eq.~\eqref{Jexp} might seem to suggest, the harmonic forms $J_i$ are typically $t^i$-dependent -- all we know is that their cohomology classes $[J_i]$ do not change with the K\"ahler class so they are allowed to vary by exact forms.\\[3mm]
On a small patch $U\subset X$, we would like to introduce the forms ${\cal J}_i$, where $i=1,\ldots ,h^{1,1}(X)$, and
\begin{equation}
 {\cal J}=\sum_i t^i{\cal J}_i \label{Jexploc}
\end{equation}
which are local $(1,1)$-forms with constant coefficients which approximate their global counterparts $J_i$ and $J$ on $U$. How are these global and local forms related?
We first note that the top forms $J\wedge J\wedge J$ and $J_i\wedge J\wedge J$ are harmonic and must, therefore be proportional
\begin{equation}
 J_i\wedge J\wedge J=c_i({\bf t})J\wedge J\wedge J\; , \label{Jrel}
\end{equation}
where $c_i({\bf t})$ are functions of the K\"ahler moduli but independent of the coordinates of $X$. By inserting Eq.~\eqref{Jexp} and integrating over $X$ we can easily compute these constants as
\begin{equation}
 c_i({\bf t})=\frac{\kappa_i}{\kappa}\; , \label{cires}
\end{equation} 
where the quantities $\kappa$ and $\kappa_i$ were defined in and around Eq.~\eqref{Vdef}. On the other hand, the relation~\eqref{Jrel} holds point-wise and, hence, has a local counterpart
\begin{equation}
 {\cal J}_i\wedge {\cal J}\wedge {\cal J}=c_i({\bf t}){\cal J}\wedge {\cal J}\wedge {\cal J}\; , \label{Jrelloc}
\end{equation}
which must involve the same constants $c_i({\bf t})$. Inserting flat forms into Eq.~\eqref{Jrelloc} then allows us to determine the  $c_i({\bf t})$ in terms of the parameters in these forms and equating these expressions to the global result~\eqref{cires} leads to constraints on the local forms ${\cal J}_i$.\\[3mm]
This global-local correspondence has an immediate implication for bundles on $X$ and their local counterparts on $U$. Consider a line bundle $L\rightarrow X$ with first Chern class $c_1(L)=k^iJ_i$ and field strength $F=-2\pi i\sum_ik^iJ_i$. Then, for the local version ${\cal F}=-2\pi i\sum_ik^i{\cal J}_i$ of the field strength we find, using Eqs.~\eqref{Jrelloc} and \eqref{cires}, that
\begin{equation}
 {\cal F}\wedge{\cal J}\wedge{\cal J}=\frac{k^i{\cal K}_i}{{\cal K}}{\cal J}\wedge{\cal J}\wedge{\cal J} \label{HYMloc}
\end{equation}
and, hence, that the local version of the Hermitian Yang-Mills equation is satisfied as long as the slope $\mu(L)=k^i{\cal K}_i$ of $L$ vanishes.\\[3mm]
To work out the above global-local correspondence more explicitly, we consider a case with two K\"ahler moduli, so $h^{1,1}(X)=2$. In this case, we can choose complex coordinates $z_a$, where $a=1,2,3$, on the patch $U\subset X$ such that
\begin{equation}
 {\cal J}_1=\frac{i}{2\pi}\sum_{a=1}^3\lambda_a dz_a\wedge d\bar{z}_a\;,\quad  {\cal J}_2=\frac{i}{2\pi}\sum_{a=1}^3dz_a\wedge d\bar{z}_a\;,\quad
 {\cal J}=\frac{i}{2\pi}\sum_{a=1}^3(\lambda_at_1+ t_2)dz_a\wedge d\bar{z}_a\; , \label{Jloc}
\end{equation} 
where the $\lambda_a$ are constants. (More specifically, starting with two arbitrary $(1,1)$ forms ${\cal J}_1$ and ${\cal J}_2$ with constant coefficients, by standard linear algebra, we can always diagonalise ${\cal J}_2$ into ``unit matrix form" and then further diagonalise ${\cal J}_1$ without affecting ${\cal J}_2$.) Inserting the above forms into Eq.~\eqref{Jrelloc} gives
\begin{equation}
 c_1({\bf t})=\frac{\sum_a\lambda_a\prod_{b\neq a}(\lambda_bt_1+t_2)}{3\prod_c(\lambda_ct_1+t_2)}\;,\quad
 c_2({\bf t})=\frac{\sum_a\prod_{b\neq a}(\lambda_bt_1+t_2)}{3\prod_c(\lambda_ct_1+t_2)} \label{c1c2res}
\end{equation} 
and equating these results to the global ones in Eq.~\eqref{cires} imposes constraints on the unknown local coefficients $\lambda_a$. However, it is not obvious that the $\lambda_a$ are K\"ahler moduli independent, particularly since the forms $J_i$ do, in general, depend on K\"ahler moduli. In the following, we will assume that this is indeed the case, although we do not, at present, have a clear-cut proof. There are two pieces of evidence which support this assumption. First, it is not obvious that equating~\eqref{c1c2res} with \eqref{cires} allows for a solution with constant $\lambda_a$ (valid for all ${\bf t}$) but we find that, in all cases which we have checked, that it does. Secondly, it is hard to see how a local calculation of the integrals in Eq.~\eqref{G1} can lead to K\"ahler moduli independent results for $\Lambda_{ijIJ}$, as four-dimensional supersymmetry demands, if the $\lambda_a$ are $t^i$-dependent. In the following, we will proceed on the assumption that the $\lambda_a$ are indeed $t^i$-independent.\\[3mm]
\subsection{An example}
To complete the above calculation we should consider a specific Calabi-Yau manifold. As before, we focus on the ambient space ${\cal A}=\mathbb{P}^1\times\mathbb{P}^3$, discussed in Section~\ref{sec:P1P3}, and use the same notation for coordinates, K\"ahler forms and K\"ahler potentials as introduced there. The Calabi-Yau hyper-surfaces $X\subset {\cal A}$ we would like to consider are then defined as the zero loci of bi-degree $(2,4)$ polynomials $p$, that is sections of the bundle $\hat{N}={\cal O}_{\cal A}(2,4)$. This manifold has Hodge numbers $h^{1,1}(X)=2$, $h^{2,1}(X)=86$ and Euler number $\eta(X)=-168$. Its second cohomology is spanned by the restrictions $\hat{J}_i|_X$, where $i=1,2$, of the two ambient space K\"ahler forms and, relative to this basis, the second Chern class of the tangent bundle is $c_2(TX)=(24,44)$. The K\"ahler class on $X$ can be parametrised by the restricted ambient space K\"ahler forms
\begin{equation}
 \hat{J}|_X=t_1\hat{J}_1|_X+t_2\hat{J}_2|_X\; ,
\end{equation}
where $t_1,t_2>0$ are the two K\"ahler parameters. Of course neither of these forms is harmonic relative to the Ricci-flat metric on $X$ associated to the class $[\hat{J}|_X]$ (as they are obtained by restricting the ambient space Fubini-Study K\"ahler forms) but there exist forms $J_i$ and $J$ in the same cohomology classes which are. In other words, $J$ and $J_i$ are the harmonic forms introduced in Eq.~\eqref{Jexp} and we demand that their cohomology classes satisfy $[J]=[\hat{J}|_X]$, $[J_i]=[\hat{J}_i|_X]$.\\[3mm]
The non-vanishing triple intersection numbers of this manifold are given by
\begin{equation}
 d_{122}=4\;,\quad d_{222}=2\quad\Rightarrow\quad  {\cal K}=d_{ijk}t^it^jt^k=2t_2^2(6t_1+t_2)\; .\label{dex}
 \end{equation}
Inserting these results into Eq.~\eqref{cires} we find
\begin{equation}
 c_1({\bf t})=\frac{2}{6t_1+t_2}\;,\qquad c_2({\bf t})=\frac{4t_1+t_2}{t_2(6t_1+t_2)}\; ,
\end{equation} 
and equating these expressions to the local results~\eqref{c1c2res} leads to the solution
\begin{equation}
 \lambda_1=6\;,\quad \lambda_2=\lambda_3=0\; ,
\end{equation}
which is unique, up to permutations of the coordinates $z_a$. This means, from Eqs.~\eqref{Jloc}, the local forms ${\cal J}_i$ and ${\cal J}$ can (after another coordinate re-scaling $z_1\rightarrow z_1/\sqrt{6}$) be written as
\begin{eqnarray}
{\cal J}_1&=&\frac{i}{2\pi}dz_1\wedge d\bar{z}_1\\
{\cal J}_2&=&\frac{i}{2\pi}\left(\frac{1}{6}dz_1\wedge d\bar{z}_1+dz_2\wedge d\bar{z}_2+dz_3\wedge d\bar{z}_3\right)\\
{\cal J}&=&\frac{i}{2\pi}\left(t_1dz_1\wedge d\bar{z}_1+t_2\left(\frac{1}{6}dz_1\wedge d\bar{z}_1+dz_2\wedge d\bar{z}_2+dz_3\wedge d\bar{z}_3\right)\right)\; . \label{Jloc1}
\end{eqnarray}
We note that ${\cal J}$ is of the form~\eqref{Jloc0} used in our local calculation and we can match expressions by setting $z_a=Z_a$ and
\begin{equation}
\beta_1=t_1+\frac{1}{6}t_2\;,\qquad \beta_2=\beta_3=t_2\; .
\end{equation}
Another interesting observation is that these forms satisfy
\begin{equation}
 {\cal J}_i\wedge {\cal J}_j\wedge {\cal J}_k=-\frac{1}{16\pi^3}d_{ijk}\bigwedge_{a=1}^3dz_a\wedge d\bar{z}_a\; , \label{isecapp}
\end{equation} 
where $d_{ijk}$ are the intersection numbers~\eqref{dex} of the manifold in question, that is, our local forms ``intersect" on the global intersection numbers. They also relate in an interesting way to the ambient space K\"ahler forms $\hat{J}_i$. So far, we have considered an arbitrary patch $U$ on $X$ but from now on let us focus on a specific choice, starting with the ambient space patch $\hat{U}\subset{\cal A}$ near the affine origin $z_\alpha\simeq 0$. This patch is of obvious interest since we know from the ambient space discussion in Section~\ref{sec:P1P3} that some wave functions localise on it. If it is sufficiently small, the defining equation of the Calabi-Yau manifold on $\hat{U}$ can be approximated by
\begin{equation}
 p=p_0+\sum_{\alpha=1}^4p_\alpha z_\alpha+{\cal O}(z^2)\; , \label{plin}
 \end{equation}
where $p_0$ and $p_\alpha$ are some of the parameters in $p$. It is possible, by linear transformations of the homogeneous coordinates on $\mathbb{P}^1$ and $\mathbb{P}^3$, to eliminate the $p_0$ term and, in the following, we assume that this has been done. Then, the Calabi-Yau manifold $X=\{p=0\}$ intersects the patch $\hat{U}$ at the affine origin and near it $X$ is approximately given by the hyper-plane equation $\sum_{\alpha=1}^4p_\alpha z_\alpha=0$. By a further linear re-definition of coordinates on the $\mathbb{P}^3$ factor of the ambient space this equation can be brought into the simpler form
\begin{equation}
 z_4=a z_1\; , \label{pdefloc}
\end{equation}
where $a$ is a constant. If we restrict the flat versions of the ambient space K\"ahler forms, as given in Eq.~\eqref{P1P3loc}, to $U$ using Eq.~\eqref{pdefloc} we find that
\begin{equation}
 \hat{J}_i|_U={\cal J}_i\; ,
\end{equation}
 provided we set $a=1/\sqrt{6}$. This means on the patch $U$ we understand the relation between ambient space K\"ahler froms $\hat{J}_i$, local K\"ahler forms ${\cal J}_i$ and their global counterparts $J_i$ on $X$.\\[3mm]
We can now extend this correspondence to (line) bundles and their connections. As in Section~\ref{sec:P1P3} we consider line bundles $\hat{L}={\cal O}_{\cal A}(k_1,k_2)$ and we restrict these to line bundles $L={\cal O}_X(k_1,k_2):=\hat{L}|_X$ on the Calabi-Yau manifold $X$. (Of course, the line bundle $L$ should be thought off as merely part of the full vector bundle of the compactification in question.) The hermitian bundle metric $\hat{H}$ for $\hat{L}$ was given in Eq.~\eqref{Hhat} and its local approximation on $\hat{U}$ in Eq.~\eqref{P1P3loc}. If we restrict this local bundle metric on $\hat{U}$ to $U$, using the defining equation~\eqref{pdefloc} with $a=1/\sqrt{6}$ we find
\begin{equation}
{\cal H}=\hat{H}|_U=\exp\left(-(k_1+k_2/6)|z_1|^2-k_2|z_2|^2-k_2|z|_3^2\right)\quad\Rightarrow\quad {\cal F}=\bar{\partial}\partial\ln{\cal H}=-2\pi i(k_1{\cal J}_1+k_2{\cal J}_2)\; .
\end{equation}
We note that this expression of ${\cal H}$ is of the general form~\eqref{Hloc0} used in the local calculation, provided we set $z_a=Z_a$ and identify
\begin{equation}
 K_1=k_1+\frac{1}{6}k_2\;,\qquad K_2=K_3=k_2\; . \label{kK}
 \end{equation}
From the discussion around Eq.~\eqref{HYMloc} we also conclude that the Hermitian Yang-Mill equation is locally satisfied for ${\cal F}$ provided that the slope $\mu(L)=d_{ijk}k^it^jt^k=2t_2(2k_1t_2+k_2(4t_1+t_2))$ vanishes. As usual, this is the case on a certain sub-locus of K\"ahler moduli space, provided that $k_1$ and $k_2$ have opposite signs.

\subsection{Wave functions and the matter field K\"ahler metric}
As the last step, we should work out the global-local correspondence for wave functions. As in Section~\ref{sec:P1P3} we consider line bundles $\hat{L}={\cal O}_{\cal A}(k_1,k_2)$ with $k_1\leq -2$ and $k_2> 0$ with a non-zero first cohomology $H^1({\cal A},\hat{L})$ whose dimension is given in Eq.~\eqref{h1A} and with harmonic basis forms $\hat{\nu}_{\hat{\bf q}}$ introduced in Eq.~\eqref{nuhat}. These line bundles restrict to line bundle $L={\cal O}_X(k_1,k_2):=\hat{L}|_X$ on the Calabi-Yau manifold $X$ with a non-vanishing first cohomology (see, for example, Ref.~\cite{Buchbinder:2016jqr})
\begin{equation}
 H^1(X,L)\cong\frac{H^1({\cal A},\hat{L})}{p(H^1({\cal A},\hat{N}^*\otimes\hat{L}))}\; . \label{H1L}
\end{equation} 
Explicit representatives for this cohomology can be obtained by restrictions $\hat{\nu}_{\hat{\bf q}}|_X$ although these forms are not necessarily harmonic with respect to any particular metric. (Also, they have to be suitably identified due to the quotient in Eq.~\eqref{H1L}. As long as $k_2<4$ the cohomology in the denominator of Eq.~\eqref{H1L} vanishes so that the quotient is trivial and the restrictions $\hat{\nu}_{\hat{\bf q}}|_X$ form a basis of $H^1(X,L)$ as stands.) Finally, we have the monomial basis $\nu_{\bf q}$ of locally harmonic forms defined in Eq.~\eqref{numon}. In summary, we are dealing with three sets of basis forms and their linear combinations, namely
\begin{equation}
\begin{array}{lllll}
 \hat{\nu}_{\hat{\bf q}}=e^{k_1|z_1|^2}\bar{z}_1^{\hat{q}_1}z_2^{\hat{q}_2}z_3^{\hat{q}_3}z_4^{\hat{q}_4}d\bar{z}_1&\quad&
 \tilde{\nu}_{\tilde{\bf q}}=e^{k_1|z_1|^2}\bar{z}_1^{\tilde{q}_1}z_2^{\tilde{q}_2}z_3^{\tilde{q}_3}z_1^{\tilde{q}_4}d\bar{z}_1&\quad&
 \nu_{\bf q}=e^{K_1|z|^2}\bar{z}_1^{q_1}z_2^{q_2}z_3^{q_3}d\bar{z}_1\\[2mm]
 \hat{\nu}(\hat{\bf a})=\sum_{\hat{\bf q}}\hat{a}_{\hat{\bf q}}\hat{\nu}_{\hat{\bf q}}&\quad&
\tilde{\nu}(\tilde{\bf a})=\sum_{\tilde{\bf q}}\tilde{a}_{\tilde{\bf q}}\tilde{\nu}_{\tilde{\bf q}}&\quad&
\nu({\bf a})=\sum_{{\bf q}}a_{{\bf q}}\nu_{{\bf q}}
\end{array}\; .
\end{equation} 
To be clear, hatted wave functions $\hat{\nu}_{\hat{\bf q}}$ are defined on the ambient space ${\cal A}$, wave functions $\tilde{\nu}_{\tilde{\bf q}}$ refer to their restrictions to the Calabi-Yau patch $U$ and the $\nu_{\bf q}$ are the harmonic wave functions on the patch $U$. \\[3mm]
Recall that we need $K_1<0$ as a necessary condition for the harmonic solutions $\nu_{\bf q}$ to have a finite norm and, by virtue of the identification~\eqref{kK}, this translates into
\begin{equation}
 K_1<0\quad\Leftrightarrow\quad -k_1>\frac{k_2}{6}\; . \label{Lcond}
\end{equation}
Hence, for this particular example, the condition $K_1<0$ is not moduli-dependent and can be satisfied by a suitable choice of line bundle.\\[3mm]  
We would like to determine the relation between the above three types of forms, or, equivalently, the relation between the coefficients $\hat{\bf a}$, $\tilde{\bf a}$ and ${\bf a}$, given that $\tilde{\nu}(\tilde{\bf a})=\hat{\nu}(\hat{\bf a})|_U$ are related by restriction and that $\tilde{\nu}(\tilde{\bf a})$ and $\nu({\bf a})$ are in the same cohomology class so must differ by a $\bar{\partial}$-exact $L$-valued $(0,1)$-form.\\[3mm]
The first of these correspondences between $\hat{\bf a}$ and $\tilde{\bf a}$ is easy to establish. Given the relation is by restriction, there is a matrix ${\cal S}$ such that $\tilde{\bf a}={\cal S}\hat{\bf a}$ and using the approximate defining equation~\eqref{pdefloc} we find that
\begin{equation}
 {\cal S}_{\tilde{\bf q},\hat{\bf p}}=\delta_{\tilde{\bf q},\hat{\bf p}}6^{\hat{q}_4/2}\; .
\end{equation} 
To establish the correspondence between ${\bf a}$ and $\tilde{\bf a}$ we first define the matrix ${\cal T}$ by
\begin{equation}
 \langle\nu_{\bf q},\tilde{\nu}_{\tilde{\bf p}}\rangle = (M{\cal T})_{{\bf q},\tilde{\bf p}} \label{Tdef}
\end{equation}
where $M$ is the local normalisation matrix computed in Eq.~\eqref{locres}. Since $\nu({\bf a})$ and $\tilde{\nu}(\tilde{\bf a})$ differ by an exact form we know that
$\langle \nu({\bf a}),\nu({\bf b})\rangle={\bf a}^\dagger M{\bf b}$ and  $\langle \nu({\bf a}),\tilde{\nu}(\tilde{\bf b})\rangle={\bf a}^\dagger M{\cal T}\tilde{\bf b}$ must be equal to each other and, since this holds for all ${\bf a}$, it follows that
\begin{equation}
 {\bf b}={\cal T}\tilde{\bf b}\; .
\end{equation} 
The explicit form of the matrix ${\cal T}$, from its definition~\eqref{Tdef}, is
\begin{equation}
 {\cal T}_{{\bf q},\tilde{\bf p}}=\delta_{q_1,\tilde{p}_1-\tilde{p}_4}\delta_{q_2,\tilde{p}_2}\delta_{q_3,\tilde{p}_3}\frac{\tilde{p}_1!\,|k_1|^{-\tilde{p}_1-1}}{q_1!\,|K_1|^{-q_1-1}}\; .
\end{equation} 
As discussed earlier, the families correspond to cohomology classes in $H^1(X,L)$ and in view of Eq.~\eqref{H1L} and  subject to possible identifications it, therefore, makes sense to label families by the hatted basis $\hat{\nu}_{\hat{\bf q}}$ on the ambient space. For simplicity of notation, we write the hated indices as ${\bf I}=\hat{\bf q}$ form now on. We also recall from Section~\ref{sec:P1P3} that these indices are non-negative and further constrained by $I_1=0,\ldots ,-k_1-2$ and $I_2+I_3+I_4\leq k_2$. With this notation, the  matter field K\"ahler metric is given by the general expression
\begin{equation}
 G_{{\bf I},{\bf J}}:=\frac{1}{2{\cal V}}({\cal S}^\dagger {\cal T}^\dagger M{\cal T}{\cal S})_{{\bf I},{\bf J}}\; .
\end{equation} 
Inserting the above results for ${\cal S}$ and ${\cal T}$ as well as the local normalisation matrix~\eqref{locres} we find explicitly,
\begin{equation}
 G_{{\bf I},{\bf J}}=\frac{{\cal N}_{{\bf I},{\bf J}}}{6t_1+t_2}\;, \label{main}
\end{equation}
where the constants ${\cal N}_{{\bf I},{\bf J}}$ are given by
\begin{equation}
 {\cal N}_{{\bf I},{\bf J}}=\frac{\pi J_1!\,I_1!\,I_2!\,I_3!\,|k_1+k_2/6|^{I_1-I_4+1}\,6^{I_4/2+J_4/2+1}}{2(I_1-I_4)!\,|k_1|^{J_1+1}k_2^{I_2+I_3+2}}\theta(I_1-I_4)\delta_{I_1-I_4,J_1-J_4}\delta_{I_2,J_2}\delta_{I_3,J_3}\; . \label{mainN}
\end{equation} 
For the lowest mode, ${\bf I}={\bf 0}$, this number specialises to
\begin{equation}
 {\cal N}_{{\bf 0},{\bf 0}}=3\pi \frac{|k_1+k_2/6|}{k_2^2}\; .
\end{equation}
A few remarks about this result are in order. First, we note that the K\"ahler moduli dependence in Eq.~\eqref{main} is in line with the result~\eqref{G1} from dimensional reduction. In general, the matter field K\"ahler metric is also a function of complex structure moduli. For our example, this dependence has dropped out completely, that is, the quantities ${\cal N}_{{\bf I},{\bf J}}$ are constants. This feature results from our linearised local approximation~\eqref{pdefloc} of the Calabi-Yau manifold, where all remaining complex structure parameters can be absorbed into coordinate re-definitions. We do expect complex structure dependence to appear at the next order, that is, if we approximate the defining equation locally by a quadric in affine coordinates. Also, our result~\eqref{main} has an implicit complex structure dependence in that its validity depends on the choice of complex structure. 
Whether neglecting the quadratic and higher terms in $z$ in Eq.~\eqref{plin} does indeed provide a good approximation depends, among other things, on the choice of coefficient in the defining equation $p$, that is, on the choice of complex structure. Another feature of our result~\eqref{main} is that it is diagonal in family space and, formally, this happens because the matrices $M$, ${\cal S}$ and ${\cal T}$ are all diagonal. We have seen in Section~\ref{sec:loc} that this is a general feature of the matrix $M$. However, ${\cal S}$ and ${\cal T}$ do not need to be diagonal in general. In our example, this happens due to the simple form~\eqref{pdefloc} of the local Calabi-Yau defining equation and the resulting diagonal form of the local K\"ahler form ${\cal J}$ in Eq.~\eqref{Jloc1}. Finally, we remind the reader that the result~\eqref{main} can only be trusted if the line bundle $L={\cal O}_X(k_1,k_2)$ satisfies the condition~\eqref{Lcond}, if the flux parameters $|k_i|$ are sufficiently large and if the family numbers ${\bf I}$ are sufficiently small, in line with our discussion in Section~\ref{sec:proj}.

\section{Conclusion}\label{sec:con}
In this note, we have reported progress on computing the matter field K\"ahler metric in heterotic Calabi-Yau compactifications. Three main results have been obtained. First, by dimensional reduction we have derived a general formula~\eqref{G1} for the matter field K\"ahler metric and we have argued that constraints from four-dimensional supersymmetry already fully determine the K\"ahler moduli dependence of this metric.\\[3mm]
Secondly, provided large flux leads to localisation of the matter field wave function, we have shown how the matter field K\"ahler metric can be obtained from a local computation on the Calabi-Yau manifold, leading to the general result~\eqref{locres}. This result, while quite general, is unfortunately of limited use, mainly since it is not expressed in terms of the global moduli of the Calabi-Yau manifold. This makes it difficult to identify the conditions for its validity and it falls short of the ultimate goal of obtaining the matter field K\"ahler metric as a function of the properly defined moduli superfields.\\[3mm]
We have attempted to address these problems by working out a global-local relationships and by expressing the local result in terms of global quantities. This has been explicitly carried out for the example of Calabi-Yau hyper-surfaces $X$ in the ambient space $\mathbb{P}^1\times\mathbb{P}^3$ but the method can be applied to other Calabi-Yau hyper-surfaces (and, possibly complete intersections) as well. Our main result in this context is the K\"ahler metric for matter fields from line bundles $L={\cal O}_X(k_1,k_2)$ on $X$ given in Eqs.~\eqref{main}, \eqref{mainN}, which is expressed as a function of the proper four-dimensional moduli fields. We have also stated the conditions for this result to be trustworthy, namely the constraint~\eqref{Lcond} on the line bundle $L$ as well as large fluxes $|k_i|$ and small family numbers. More details and examples will be given in a forthcoming paper.\\[3mm]
The global-local relationship established in this way points to two problems of localised calculations both of which are intuitively plausible. First, the large flux values demanded by localisation typically also lead to large numbers of families. For this reason, there is a tension between localisation and the phenomenological requirement of three families. Secondly, large flux typically leads to a ``large" second Chern class $c_2(V)$ of the vector bundle which might violate the anomaly constraint $c_2(V)\leq c_2(TX)$. Hence, there is also a tension between localisation and consistency of the models. It remains to be seen and is a matter of ongoing research whether consistent three-family models with localisation of all relevant matter fields can be constructed.\\[3mm]
It is likely that some of our methods can be applied to F-theory and be used to express local F-theory results in terms of global moduli of the underlying four-fold. It would be interesting to carry this out explicitly and check if the tension between localisation on the one hand and the phenomenological requirement of three families and cancelation of anomalies on the other hand persists in the F-theory context.\\[5mm]
%
{\bf Acknowledgements}  A.L. is partially supported by the EPSRC network grant EP/N007158/1 and by the STFC grant~ST/L000474/1.
E.~I.~B. and A.~C.~would like to thank the Department of Physics of the University of Oxford where some of this work has been carried out for warm hospitality.

\newpage
\bibliography{bibfile}{}
\bibliographystyle{utcaps}
\end{document}